\documentclass[ reprint, superscriptaddress, groupedaddress, showpacs, amsmath,amssymb, aps,pre]{revtex4-1} % {{{

\usepackage{graphicx}% Include figure files
\usepackage{dcolumn}% Align table columns on decimal point
\usepackage{bm}% bold math
\usepackage{color}
\usepackage{hyperref}% add hypertext capabilities
\graphicspath{{images/},{figures/}}
\usepackage[utf8]{inputenc}
\def\v#1{{\bf#1}}
\def\be{\begin{equation}}
\def\ee{\end{equation}}
\def\bea{\begin{eqnarray}}
\def\eea{\end{eqnarray}}

\newcommand{\bfalpha}{\mbox{\boldmath$\alpha$\unboldmath}}

\newcommand{\bfsigma}{\mbox{\boldmath$\sigma$\unboldmath}}

\newcommand{\zitt}{{\rm zitt}}

\def\ncal{\mbox{$\cal N\,$}}
\def\<{\langle}
\def\>{\rangle}

\begin{document}

\newcommand{\ifunamint}{Instituto de F\'{\i}sica, Universidad Nacional Aut\'onoma de M\'exico, 01000 M\'exico D.F., Mexico}
\title{Stern-Gerlach splitters for lattice quasispin}% Force line breaks with \\
% \title{Stern-Gerlach splitters based on quasispin}% Force line breaks with \\

\author{A. S. Rosado$^{2}$}
\author{J. A. Franco-Villafa\~ne$^1$} % \email{xxxXX@ifuap.buap.mx}
\author{C. Pineda$^2$}
\author{E. Sadurn\'i$^1$} \email{sadurni@ifuap.buap.mx}

\affiliation{$^1$Instituto de F\'isica, Benem\'erita Universidad Aut\'onoma de Puebla,
Apartado Postal J-48, 72570 Puebla, M\'exico \\ $^2$\ifunamint}

\date{\today}% It is always \today, today,

\begin{abstract}
We design a Stern-Gerlach apparatus that separates quasispin components on the lattice, without the use of external fields. The effect is engineered using intrinsic parameters, such as hopping amplitudes and on-site potentials. A theoretical description of the apparatus relying on a generalized Foldy-Wouthuysen transformation beyond Dirac points is given. Our results are verified numerically by means of wave-packet evolution, including an analysis of \textit{Zitterbewegung} on the lattice. The necessary tools for microwave realizations, such as complex hopping amplitudes and chiral effects, are simulated.
\end{abstract}



\maketitle

%\tableofcontents
% }}}
\section{Introduction \label{sec:1}} % {{{

%> - Decir que la simulaciones son importantes
%> - Mencionar simulaciones en el campo de ecuación de Dirac y las microondas como simuladores
%> - Relacionar tantito con grafeno y otras estructuras.
%> - Resaltar aspectos como la medición y la separacion de la funcion de onda
%> - y lo del cambio de signo en las microondas.
%"edge states and Majorana fermions"
%
%
%Relacionar los quantum analogs, interfaces entre diferentes sistemas fisicos, 
%la posibilidad de usar en un sistema otros sistemas.
%Tomar lo mejor de uno y de otro. 
%
%Mencionar aca tanto la ecuación de Dirac, por su interes teórico
%como para profundizar en el entendimiento de la naturaleza, 

Quantum emulations have been increasingly important 
for theorists and experimentalists in areas such as ultracold atoms \cite{oberthaler2006, bloch2005,
oberthaler1996, esslinger2013, struck2012}, quantum and microwave billiards \cite{kuhl2010, barkhofen2013, bittner2010, bellec2013}, plasmonic circuits~\cite{2014NatSR...4E7314M}, and artificial solids in general \cite{polini2013, gomes2012}. 
The concept can be used to engineer quantum dynamics not readily accessible
in naturally occurring physical systems, e.g., elementary particles or charge carriers in solids \cite{semenoff1984, geim2007}. 
For some years, the effective Dirac theories emerging in honeycomb lattices
and linear chains \cite{castroneto2009, roati2007, inguscio2007, franco2013, sadurni2013, sadurni2010} have led researchers to consider the use of quasispin as
an internal degree of freedom capable of supporting the long-pursued
realization of qubits in solid-state physics. This interesting degree of
freedom has the property of being nonlocal, inherent to the
crystalline structure, and sufficiently robust as to provide upper and lower
bands around conical (Dirac) points in the spectrum. In the same context, there has been a recent interest
in Majorana fermions \cite{mancini2015, wilczek2009, alicea2012}, as their topological nature may provide
robustness with respect to decoherence, hence increasing the life 
of qubits, and thus extending the reach of potential applications. 
Several theoretical developments take advantage of
quasispin \cite{kagalovsky1999, bernard2001} and some experiments in lattices have observed their effects,
e.g., \textit{Zitterbewegung} in photonic structures \cite{Dreisow2010}.

But how does one measure quasispin on the lattice? One of the goals of this paper is to gain access to this degree of freedom by designing an
interaction on bipartite lattices with the following features: (a) an adjustable
coupling with particles' quasispin, (b) a localized region where the
interaction occurs, and (c) an intrinsic generation of the interaction using
lattice parameters. It is worth mentioning that the electron's true spin is not easily accessible when immersed in a solid \cite{meier2007}.

Our tasks demand an exploration of tight-binding models, oriented to an experimental
setup in microwave resonators. We establish the realization of Dirac's equation in
a one-dimensional setting and solve the problem of how to split the two components of the wave function, 
namely particle-antiparticle components, or, in the language of solid-state physics, the upper and lower bands. Under these 
circumstances, and using the Foldy-Wouthuysen (FW) transformation, 
we design and test a spatially localized Stern-Gerlach 
splitter represented by a banded matrix, to be used in the context of Dirac-like dynamics. 
In this case, 
the experimental restrictions imposed by most realizations come in the form of short-range interactions. We provide a successful geometric proposal in compliance with such restrictions, using microwave resonators coupled by proximity. 

%As we know well, electrons' true
%spin can be probed by means of external magnetic fields. For charged particles
%moving in solids, the coupling to such a field is so small that the most
%prominent effects come in the form of Landau levels, rather than spin
%precession. Other visible effects in theoretical calculations involve Rashba
%couplings \cite{}, but to our knowledge, such couplings cannot be controlled
%externally in a simple way, let alone the actual design of a polarizer using
%this type of interactions (i.e. a Stern-Gerlach apparatus plus a shutter). 

%???
%Taking the advantages of one particular
%realization into another, and using what has been learned in one field
%into the other is a cornerstone in physics. This is particularly built in
%quantum mechanical realizations, as the underlying mathematical 
%structure is simply Hilbert space. 
%???

%Here we see that simulating across fields (quantum simulation) is indeed rewarding.
%It has been shown that  
%
%the last years, several combinations have been of particular interest
%for the community, starting with [ ], and [ ]. In the latest years 
%the study of Dirac equation has been given some attention. Graphene
%has provided a unique opportunity to fine tune the parameters 
%of, with todays technologies, otherwise unaccessible regimes of
%massless Dirac fermions. 

%Additional efforts in similar directions have led to the first realization of Majorana fermions in solid state physics, where
%a typical identification of Dirac q-bits comes from a pair of Majoranas \cite{mancini2015, wilczek2009, alicea2012}.

We approach the problem in three different stages. First, in Sec. \ref{sec:2}, we study the lattice structure using full-band Dirac equations
\cite{sadurni2010} and provide a 
% method for designing interactions by means of a 
generalized FW transformation in Sec. \ref{sec:2A}. The explicit construction of
the beam splitter as a potential is achieved in Sec. \ref{sec:2B}. In Sec.
\ref{sec:3}, we study wave-packet dynamics using numerical simulations with two
important results: in Sec. \ref{sec:3A}, we show that unpolarized beams exhibit
\textit{Zitterbewegung}, while in Secs. \ref{sec:3B} and \ref{sec:3C}, we test the splitter
efficiency. With the aim of ensuring the feasibility of our model, in Sec.
\ref{sec:4} we establish the robustness of the system under random
perturbations of parameters. Our study is applicable to any tight-binding (TB)
array with the aforementioned structure, but, as a final step, in Sec.
\ref{sec:5} we focus on plausible experiments in microwave cavities. Section
\ref{sec:5A} describes the necessary specifications for the implementation and
Sec. \ref{sec:5B} gives an explicit construction that produces negative couplings
and level inversion. We conclude in Sec. \ref{sec:6}.

% }}}
\section{Intrinsic Stern-Gerlach apparatus  \label{sec:2}} % {{{
\subsection{Quasispin and generalized FW transformations  \label{sec:2A}} % {{{

\begin{figure}[t] % {{{
\includegraphics{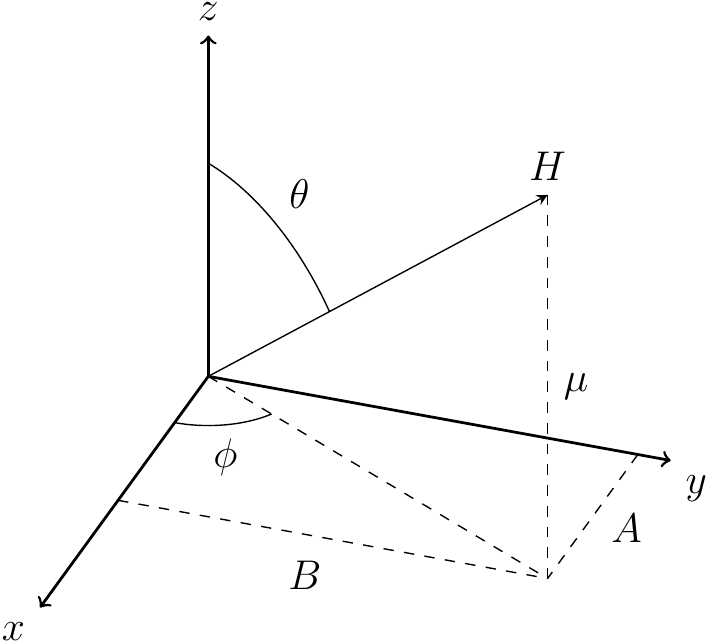} 
\caption{\label{fig:-1}
A visualization of the FW transformation. 
% Using a rotation along
% the $y$ by an angle $\theta$, followed by a rotation around the $z$ axis by 
% an angle $\phi$ 
Using a rotation around the $z$ axis by 
an angle $\phi$, followed by a rotation around 
$y$ by an angle $\theta$,
would rotate the eigenstates of $\sigma_z$ to the eigenstates
of the Hamiltonian given by Eq.~(\ref{1}). In this visualization, the angles are considered
to be scalars since they are operators that commute with the Hamiltonian. 
% Space holder for Foldy Wouthuysen transformation
}
\end{figure} % }}}
Let us define our periodic system, with the aim of generalizing the usual FW unitary rotation \cite{foldy1950, devries1970}.
Consider a one-dimensional lattice, with sites characterized by the positions
$n\in \mathbb Z$, and position basis $\{ |n\> \}_{n\in \mathbb Z}$. 
We deal with a typical TB model in this setting,
with hopping parameter $\Delta$ and potential $V$,
\begin{align}
H &= \Delta T + \Delta T^{\dagger} + V \nonumber \\ 
&= \sum_{n=-\infty}^{\infty} \Delta |n\>\<n+1| + \mbox{h.c.} +V_n|n\>\<n|
\label{1}
\end{align}
where the translation operator is defined via $T|n\> = |n+1\>$ and a position-dependent potential $V = \sum_n V_n |n\> \<n|$ has been introduced.  We have shown \cite{sadurni2010} that this
Hamiltonian can be written in Dirac form without approximations, with suitable
definitions of  Dirac matrices $\bfalpha$ in terms of projectors onto even and
odd site numbers, 
\begin{align}
H= \Delta \bfalpha \cdot \v \Pi + V
\label{3}
\end{align}
with the kinetic operators
\begin{align}
\Pi_1 &\equiv 1+ \frac{1}{2} \sum_{n} |n-2\>\<n | +  |n\>\<n-2 | 
 =1+ \frac{T^2+ (T^{\dagger})^2 }{2} \nonumber \\
\Pi_2 &\equiv \frac{i}{2} \sum_{n} |n-2\>\<n | -  |n\>\<n-2| 
=\frac{T^2 -(T^{\dagger})^2}{2i}
% \nonumber
\label{eq:pis}
\end{align}
and the Dirac matrices 
\begin{align}
\alpha_1 \equiv  \sum_{n \, \mbox{\scriptsize even}} |n+1\>\<n | +  |n\>\<n+1 | \nonumber \\
\alpha_2 \equiv  i \sum_{n \, \mbox{\scriptsize even}} |n+1\>\<n | -  |n\>\<n+1 |
\label{5}
\end{align}
satisfying the usual conditions, as proved in \cite{sadurni2010}.

Bipartite lattices with alternating on-site potential energies $E_1$, $E_2$ entail the use of the
potential
\begin{equation}
V = E_0 + \mu \beta,
\end{equation}
where the average energy $E_0 =(E_1 + E_2)/2$ and splitting
%$\mu =(E_2-E_1)/2$ are used. (version original)
$\mu =(E_1-E_2)/2$ are used. 
Additionally, we have considered the operator $\beta$, here defined as 
\begin{equation}
%\beta \equiv \sum_{n \, \mbox{\scriptsize even}} |n+1\>\<n+1 | -  |n\>\<n |,(version original)
\beta \equiv \sum_{n \, \mbox{\scriptsize even}} |n\>\<n |- |n+1\>\<n+1 |.
\label{6}
\end{equation}
Our lattice operators (\ref{5}) and (\ref{6}) satisfy the relations
$\left\{ \alpha_i, \beta \right\}=0, \left\{\alpha_i, \alpha_j \right\}=2\delta_{ij}, \left[ \alpha_1, \alpha_2\right]=2 i \beta$.
This reordering of our original TB Hamiltonian leads to 
an effective Dirac Hamiltonian of the form
\begin{equation}
H = \Delta \bfalpha \cdot \v \Pi + \mu \beta + E_0.
\label{7}
\end{equation}
The spectrum of $H$ is $E_{k,\pm} = E_0 \pm \sqrt{4\Delta^2 \cos^2
k + \mu^2}\equiv E_0 \pm E_k$ and, most importantly, its eigenfunctions are
written as spinors with up and down components represented by amplitudes in the even and odd sublattices. 
Here we remark that this spinorial form of the eigenfunctions and, in general, of any wave packet on the lattice is in itself an additional discrete degree of freedom, and thus gives rise to the name: quasispin. As previously noted, quasispin is entirely nonlocal, given that it is a direct manifestation of the bipartite nature of the lattice.\\
Returning to the discussion, we have the following complete set of eigenfunctions
\begin{equation}
\<n|k,s \> = e^{ikn} \left(\begin{array}{c}  u^{+}_{k,s}  \\
u^{-}_{k,s}   \end{array}\right), \, u^{\pm}_{k, s} = 
s^{\pm 1/2} \sqrt{\frac{E_k \pm s \mu}{4\pi E_k }},
\label{8}
\end{equation}
where $n$ is an even index, $k$ is the wave number in the reduced Brillouin zone $0 <k<\pi $, and
$s =\pm$ is the index of upper and lower bands.  
For the latter use, we introduce the parameter $\kappa$ around the conical
point $k= \pi/2 - \kappa/2$. This yields the following eigenvalues $p_i$ of $\Pi_i$: 
\begin{align}
p_1 \approx -\frac{\kappa^2}{2}, \qquad p_2 \approx \kappa
\label{9}
\end{align}
for momenta near the conical point. This shows that $p_2$ survives,
playing the role of an effective momentum of a one-dimensional (1D) Dirac equation. 

In order to show the role of quasispin in the solutions, one can solve the eigenvalue problem without any
approximation by means of a rotation in the space $(\alpha_1, \alpha_2,
\beta)$. This is the FW transformation explained in Fig. \ref{fig:-1}, 
which maps the site model (even/odd sites) to a qubit system of positive
and negative energies \cite{sadurni2013, sadurni2010}. 
In terms of Pauli matrices, we write $\alpha_1=\sigma_1,
\alpha_2=\sigma_2, \beta=\sigma_3$ and we define a vector $\v v$ with
components $v_1 = \Delta \Pi_1, v_2 = \Delta \Pi_2, v_3=\mu$.
With these
definitions, $H$ becomes a pure spin-orbit interaction,
\begin{equation}
H-E_0 = \v v \cdot \bfsigma, \qquad 
  \left[ v_i,v_j \right] = \left[ v_i,\sigma_j \right] = 0.
\label{10}
\end{equation}
This allows one to rotate the vector $\v v$ independently of $\sigma$, with the aim
of making it parallel to $\v z$. Equivalently, the rotation is
represented by a unitary transformation $U_{\mbox{\scriptsize FW}}$
which block diagonalizes $H$, 
\begin{equation}
%U_{\mbox{\scriptsize FW}} = \exp \left( \frac{i \theta}{2} \sigma_2 \right) \exp \left( \frac{i \phi}{2} \sigma_3 \right)(version original)
U_{\mbox{\scriptsize FW}} = \exp \left( -\frac{i \phi}{2} \sigma_3 \right) \exp \left(- \frac{i \theta}{2} \sigma_2 \right)
\label{eq:ufw}
\end{equation}
 In our case, this rotation allows us to guide the design of the
 polarizer. The exponential is understood in terms of trigonometric
functions, where the angles are operators defined by
% 
% 
% where the angles $\theta$ and $\phi$ are to be understood as operators, leading to 
% the relations
% 
% 
% 
% The angles $\theta, \phi$ are functions of commuting operators $\Pi$ or,
% equivalently, functions of $T, T^{\dagger}$:
\begin{align}
\sin \theta &= \frac{\Delta (T+T^{\dagger})}{\sqrt{\Delta^2(T+T^{\dagger})^2+\mu^2}}, \nonumber \\ 
\cos \theta &= \frac{\mu}{\sqrt{\Delta^2(T+T^{\dagger})^2+\mu^2}}
\label{eq:thetarelations}
\end{align}
and
\begin{align}
\cos \phi = \frac{1}{2}(T+T^{\dagger}), \quad \sin \phi = \frac{1}{2i}(T-T^{\dagger}).
\label{eq:phirelations}
\end{align}
Formula (\ref{eq:ufw}) involves trigonometric functions of half angles, so we
provide their expressions for completeness [we note here that $(H-E_0)^2$ is
independent of Pauli matrices],
\begin{align}
\cos \left( \frac{ \theta}{2} \right) &= \sqrt{ \frac{\sqrt{(H-E_0)^2} + \mu}{2 \sqrt{(H-E_0)^2}} }, \nonumber \\
\sin  \left( \frac{ \theta}{2} \right) &= \sqrt{ \frac{\sqrt{(H-E_0)^2} - \mu}{2 \sqrt{(H-E_0)^2}} }, 
\label{14}
\end{align}
and
\begin{align}
\cos \left( \frac{\phi}{2} \right) &= \frac{1}{2}(T^{1/2}+(T^{\dagger})^{1/2}), \nonumber \\ 
\sin \left( \frac{\phi}{2} \right) &= \frac{1}{2i}(T^{1/2}-(T^{\dagger})^{1/2}).
\label{15}
\end{align}
With the unitary operator $U_{\mbox{\scriptsize FW}}$, the transformation yields, in a very clean way,
% \begin{multline*}
\begin{align*}
H_{\mbox{\scriptsize FW}} 
&= U_{\mbox{\scriptsize FW}}^{\dagger} H U_{\mbox{\scriptsize FW}} \\ 
&=\begin{pmatrix} 
E_0 +
% \sqrt{\Delta^2(T+T^{\dagger})^2+\mu^2}
\sqrt{(H-E_0)^2}  
& 0   \\  
0 & E_0 - 
% \sqrt{\Delta^2(T+T^{\dagger})^2+\mu^2}
\sqrt{(H-E_0)^2}  
\end{pmatrix} 
\end{align*}
% \end{multline*}
where
$\sqrt{(H-E_0)^2}=\sqrt{\Delta^2(T+T^{\dagger})^2+\mu^2}$.

Adding the next-to-nearest-neighbor interaction in Eq. ~(\ref{1}) would require a
modification of the definitions~(\ref{eq:pis}). However, the
program of the present section could also be carried out in a very similar fashion.
The addition of the quartic translational terms in Eq.~(\ref{eq:pis}) would
change Eqs.~(\ref{eq:thetarelations}) and (\ref{eq:phirelations}), and would thus make the propagation in the two
bands slightly different. A splitter could thus also be designed, but an
asymmetry in the two components would indeed show up in the asymptotic
evolution. 
% 
% {\cp Creo que decir $\sqrt{(H-E_0)^2}$ puede resultar confuso, pues uno no mas podria decir $\sqrt{(H-E_0)^2} = H-E_0$. Mejorar.}
% {\bf A.R. Podemos utilizar en cambio
% $\sqrt{(H-E_0)^2}=\sqrt{\Delta^2(T+T^{\dagger})^2+\mu^2}$ }
%
% }}}
\begin{figure} % {{{
\includegraphics{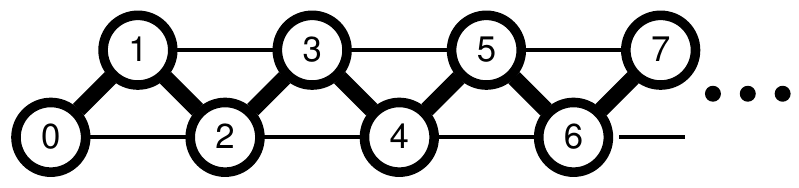} \\
\includegraphics{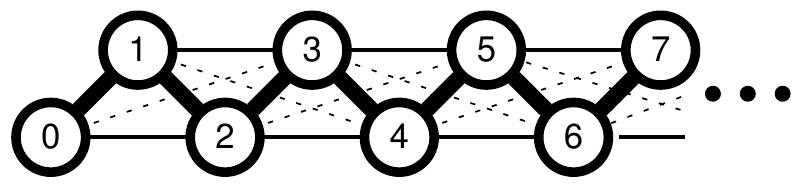}
\caption{\label{fig:-0.1} Lattice topologies corresponding to the polarizer, up to second neighbors (top)
and third neighbors (bottom). Thick lines correspond to the (strongest) nearest-neighbor interaction, thin lines are the next-to-nearest-neighbor interactions, and,
finally, dashed lines are the weakest (and in the top model neglected) third-nearest-neighbor
interactions.}
\end{figure} % }}}
\subsection{The Stern-Gerlach apparatus as an interaction \label{sec:2B}} % {{{

Now that we have derived a block-diagonal Hamiltonian, we are in the position
to introduce an interaction which couples differently with positive- and
negative-energy solutions. Moreover, we shall see that the range of such
interaction can be controlled at pleasure. A diagram is shown in Fig. \ref{fig:-0.1}.

In classical relativistic dynamics, the double sign of the kinetic energy could be used to produce two types of behavior in the presence of a potential well. If $V(x)$ interacts attractively for positive solutions (charges), the opposite case will be a potential barrier acting on negative solutions (holes):
\begin{equation}
E= \pm \sqrt{c^2p^2 + m^2 c^4}  + V(x).
\label{17}
\end{equation}
Thus, one type of solution would be allowed to enter in a certain region while the other would be rejected; we may regard $V(x)$ as a gate keeper. We must note, however, that quantum dynamics gives rise to interference phenomena producing transmission and reflection in both of the aforementioned situations. The simplest way to separate both types of waves is by introducing a potential of the type
\begin{equation}
V_{\pm}(x) = \begin{cases}
V(x) & \text{for particles} \\
0 & \text{for holes}
\end{cases}
\label{17.1}
\end{equation}
Since the FW transformation does the job of decoupling both types of solutions, we introduce at the level of $H_{\mbox{\scriptsize FW}}$ a potential $V_{\mbox{\scriptsize FW}}$ that separates the components as in (\ref{17.1}),
\begin{equation}
\tilde H_{\mbox{\scriptsize FW}} = 
H_{\mbox{\scriptsize FW}} + \frac{\v 1 + \sigma_3}{2} \otimes  V_{\mbox{\scriptsize FW}}
\label{18}
\end{equation}
or in matrix form,
\begin{equation}
\begin{pmatrix}
E_0 + V_{\mbox{\scriptsize FW}} + \sqrt{(H-E_0)^2}  & 0   \\ 
0 & E_0 - \sqrt{(H-E_0)^2} 
\end{pmatrix}.\nonumber
\label{19}
\end{equation}
In order to find the true potential $V$ operating at the level of lattice sites
and neighbor couplings, we must return to our original description by means of
the inverse FW transformation,
\begin{equation}
V(N) = U_{\mbox{\scriptsize FW}} \, V_{\mbox{\scriptsize FW}} \, U^{\dagger}_{\mbox{\scriptsize FW}}.
\label{20}
\end{equation}
Direct computations lead to a $2 \times 2$ block form of $V$. For instance,
% \begin{multline}
% V_{11} = e^{-i\phi/2}\cos\left(\frac{\theta}{2} \right) V_{\mbox{\scriptsize FW}} \cos\left(\frac{\theta}{2} \right) e^{i\phi/2} \\
% + e^{-i\phi/2}\sin\left(\frac{\theta}{2} \right) V_{\mbox{\scriptsize FW}} \sin\left(\frac{\theta}{2} \right) e^{i\phi/2}.
% \end{multline}
\begin{multline}
V_{11} = e^{-i\phi/2}\cos\left(\frac{\theta}{2} \right) V_{\mbox{\scriptsize FW}} \cos\left(\frac{\theta}{2} \right) e^{i\phi/2}
\label{v11}
%\label{21}
\end{multline}
Here we may choose $V_{\mbox{\scriptsize FW}}$ at will, but using site number
kets makes it easier to provide locality: $\<n| V_{\mbox{\scriptsize FW}}  |n'
\> = \delta_{n,n'} V_{\mbox{\scriptsize FW}}(n) $. 
The site dependence of $V$ can be obtained by inserting a complete set of Bloch waves. Let us define
\begin{equation}
I_{n}^{s, s'} \left(\frac{\mu}{\Delta}\right) 
  \equiv \int_{-\pi}^{\pi} dk \sqrt{\frac{E_k + s \mu}{E_k}} \, e^{ik(n-s'/2)}, 
\label{22}
\end{equation}
 with $s,s' = \pm$ and  $n \in \mathbf{Z}$.
The potential blocks are then
% \begin{multline}
% \<n|V_{11}|n'\> = 
%   \frac{1}{8\pi^2} \sum_{m=-\infty}^{\infty} V_{\mbox{\scriptsize FW}}(m) \\ 
%   \times 
%   \left[ I_{n'-m}^{++} \left( I_{n-m}^{++}\right)^* 
%               + I_{n'-m}^{--} \left( I_{n-m}^{--}\right)^*  
%   \right],
% \label{23}
% \end{multline}
\begin{multline}
\<n|V_{11}|n'\> = 
  \frac{1}{8\pi^2} \sum_{m=-\infty}^{\infty} V_{\mbox{\scriptsize FW}}(m)  
%   \times 
   I_{n'-m}^{++} \left( I_{n-m}^{++}\right)^*  
\label{23}
\end{multline}
for even $n$ and $n'$, 
% \begin{multline}
% \<n|V_{21}|n'\> = 
%   \frac{1}{8\pi^2} \sum_{m=-\infty}^{\infty} V_{\mbox{\scriptsize FW}}(m) \\ 
%   \times 
%   \left[ I_{n'-m}^{--} \left( I_{n-m}^{+-}\right)^* 
%       - I_{n'-m}^{++} \left( I_{n-m}^{-+}\right)^*  \right], 
% \label{24}
% \end{multline}
\begin{multline}
\<n|V_{21}|n'\> = 
  \frac{1}{8\pi^2} \sum_{m=-\infty}^{\infty} V_{\mbox{\scriptsize FW}}(m)  
% \\  \times 
  I_{n'-m}^{-+} \left( I_{n-m}^{+-}\right)^*
\label{24}
\end{multline}
for even $n$ and odd $n'$, and finally
% \begin{multline}
% \<n|V_{22}|n'\> = 
%   \frac{1}{8\pi^2} \sum_{m=-\infty}^{\infty} V_{\mbox{\scriptsize FW}}(m) \\ 
%   \times 
%   \left[ I_{n'-m}^{-+} \left( I_{n-m}^{-+}\right)^* 
%   + I_{n'-m}^{+-} \left( I_{n-m}^{+-}\right)^*  \right], 
% \label{eq:blocks:odd:odd}
% \end{multline}
\begin{multline}
\<n|V_{22}|n'\> = 
  \frac{1}{8\pi^2} \sum_{m=-\infty}^{\infty} V_{\mbox{\scriptsize FW}}(m) 
%   \\ \times 
 I_{n'-m}^{-+} \left( I_{n-m}^{-+}\right)^* 
\label{eq:blocks:odd:odd}
\end{multline}
for odd $n$ and $n'$. 
It is advantageous to write our result in the form of the series over
$m$ above: when the range of $V_{\mbox{\scriptsize FW}}$ is limited, the
summation over $m$ involves only a few terms. In the extreme case of a
pointlike gate keeper in the FW picture, $m=0$ is the only contribution in
$V$. Moreover, the limits $\mu \gg \Delta$ and $\mu \ll \Delta$ provide useful
approximations,
\begin{multline}
I_{n}^{s,s'} \approx \frac{4\sqrt{1+s}s'(-)^{n}}{s'-2n} + O \left(\frac{\Delta}{\mu}\right)
\end{multline}
and in the opposite regime,
\begin{equation}
I_{n}^{s,s'} \approx \frac{4s'(-)^{n}}{s'-2n} + O \left(\frac{\mu}{\Delta} \right).
\label{27}
\end{equation}

\begin{figure} % {{{polarizermatrix
\includegraphics[width=8.6cm]{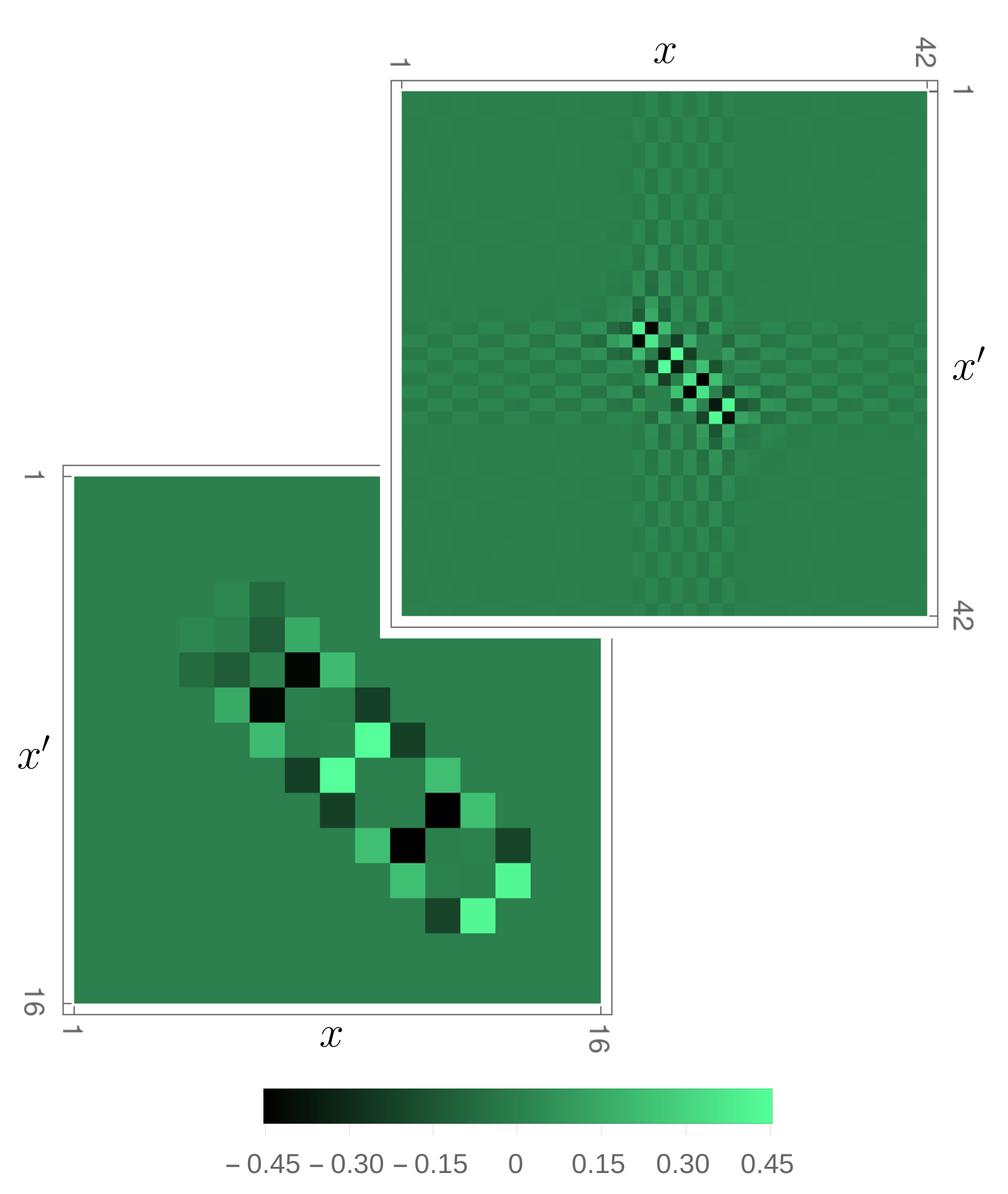} 
\caption{\label{polarizermatrix} Top: Matrix form of the nonlocal complete polarizer
potential. The interaction zone contains different
on-site energies indicated by the alternating pixel intensities in the diagonal. Bottom: Matrix form of a geometrical polarizer potential with
range $\rho=10$ and no on-site potential. Only couplings to first and second
order have been included. Both potentials are given in units of $\Delta$.}
\end{figure}

According to (\ref{23})-(\ref{eq:blocks:odd:odd}), these expansions
show that the resulting potentials in space are represented by banded matrices,
which we proceed to display as densities without approximations in Fig. \ref{polarizermatrix}.
The numerical evaluation of matrix elements shows that a finite number of neighbors is a reasonable approximation. For second- and third-nearest-neighbor interactions, we depict the resulting localized arrays in Fig. \ref{fig:-0.1}.

 % }}}

% }}}
% }}}
\section{Dynamical study  \label{sec:3}} % {{{
In this section, we shall study two different phenomena. The first is 
a ``free-particle'' effect: \textit{Zitterbewegung}. 
% Propuesta de Tony de Zitter
Since its proposal by Schrödinger, \textit{Zitterbewegung} has been understood as a
rapid oscillatory motion that is a product of the interference between positive- and
negative-energy states present in the initial composition of a Dirac spinor.
For this oscillatory phenomena to be observed, these positive- and negative-energy states must have a sufficiently large overlap in position space.  This
has, in fact, been emulated in other experimental realizations of the Dirac
equation~\cite{rusin2007,Dreisow2010,gerritsma2010}. 
In this work, we develop a clean derivation that will allow us
to make a stationary phase approximation leading to a $\sqrt{1/t}$ decay
of the oscillatory part of the amplitude. In addition, we shall consider the effect of a designed potential that can spatially
separate efficiently a function in its ``big'' and ``small'' contribution. The efficiency of the splitter shall be characterized by means of reflection and transmission coefficients for each spin component.

\subsection{Wave packet dynamics  \label{sec:3A}} % {{{

\textit{Zitterbewegung} is the hallmark of unpolarized beams. Effective relativistic
wave equations produce oscillatory phenomena in the evolution of
single-component spinors on the lattice \cite{Dreisow2010}. At the heart of this effect
lies the FW picture and the corresponding rotated quasispin: an observable
associated to upper and lower energy bands. The outcome of the evolution will
be a superposition of "particles" and "antiparticles" as long as the initial
condition is a mixture of such quantum number. An obvious implication is
that \textit{Zitterbewegung} should be present in any theory with binary lattices.
Noteworthy is the fact that the approximation of Bloch momenta around Dirac
points is not the essential ingredient; we may find \textit{Zitterbewegung} in
situations where the initial wave packet is a superposition of all energies in
both bands, with non-negligible momentum components. We proceed to analyze such
physical situations. 

\begin{figure} % {{{zittercompleta
\includegraphics{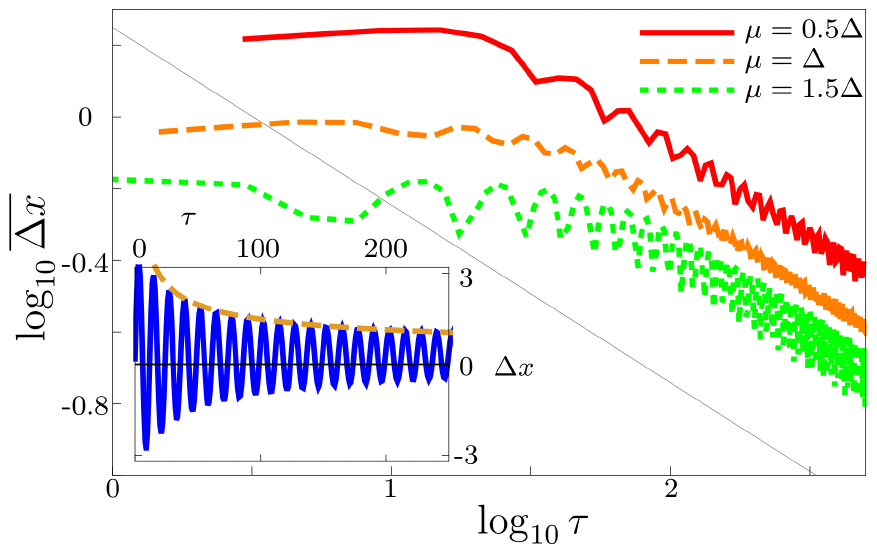} 
\caption{\label{zittercompleta} \textit{Zitterbewegung} of the wave packet. On the left panel, we
see the oscillations of $x_\zitt$ without ballistic motion, as well as the decay of
amplitude predicted by stationary phase approximations. On the right panel, we
see the same rate of amplitude decay for three different effective masses, $\mu$.
$\overline{\Delta x}$ is the averaged maximum amplitude of the oscillations,
while $\tau=t/T_\chi$ and $T_\chi$ is the characteristic time of the simulation given by $T_\chi=\hbar/\Delta$.}
\end{figure}

Setting $\hbar=1$, we define the initial wave packet as 
\begin{equation}
|\psi_0\>= \int_{0}^{\pi} dk \sum_{s=\pm} \psi_{k,s} |k, s\> 
         = \sum_{n} \psi_n |n\>.
\label{28}
\end{equation}
We are interested in the average position at time $t$. In order to recover the
usual definition of position $x$ and momentum $p=-i \partial/\partial
x$ in the continuous limit, we work with position operators defined over dimers
(pairs of sites) and lattice constant $a$,
\begin{equation}
X=\frac{a}{2} \sum_{n\, \mbox{\scriptsize even}} n \left[ |n\>\<n| + |n+1\>\<n+1|\right]
\label{29}
\end{equation}
with the property
\begin{equation}
\left[ T^2, X \right] = - a T^2, \qquad \left[ X, \sigma_{\pm} \right] =0
\label{30}
\end{equation}
(note though that it is the operator $T^2$ and not $T$ that satisfies this
property). In the Heisenberg picture, we obtain
% and working in the Heisenberg picture
\begin{equation}
\dot X = a \sigma_2, \qquad \dot{\v \Pi} = 0,
\label{31}
\end{equation}
which leads to 
\begin{multline}
X(t) = X(0) - 2 a t \frac{\Delta \Pi_2}{H}  \\
             + a \left[ \sigma_2(0) - 2 \frac{\Delta \Pi_2}{H} \right] 
                      \int_{0}^{t} dt \, e^{-2itH}.
\label{32}
\end{multline}
The first two terms describe the usual classical dynamics for a free particle, 
while the oscillations (i.e., the \textit{Zitterbewegung}) come from the third term. 
The relevant part of the expectation value with respect to the state $|\psi_0\>$ is
thus
\begin{equation}
x_\zitt \equiv \bigg \< \left[ \sigma_2(0) - 2 \frac{\Delta \Pi_2}{H} \right] 
            \int_{0}^{t} dt \, e^{-2itH} \bigg \>_{\psi}.
\label{33}
\end{equation}
After inserting energy kets (\ref{8}) and performing the time integral, we 
can write 
\begin{equation}
x_\zitt = \sum_{s, s'} J_{s, s'} + \sum_{s} I_{s}
\label{34}
\end{equation}
where $I$ and $J$ are Bloch-momentum integrals of the type
\begin{multline}
J_{s,s'} \equiv 
 \int_{0}^{\pi} dk 
   \frac{e^{-iE_{k,s}t} \sin(E_{k,s}t) }{E_{k,s}}
   \psi_{k,s} \psi^*_{k,s'} \\ 
   \times i 
   \left[ 
     u^+_{k,s} (u^{- }_{k,s'})^* - u^-_{k,s} (u^{+ }_{k,s'})^* 
   \right]
\label{35}
\end{multline}
and 
\begin{equation}
I_{s} \equiv \int_{0}^{\pi} dk \frac{e^{-iE_{k,s}t} \sin(E_{k,s}t) \sin k}{(E_{k,s})^2}|\psi_{k,s}|^2. 
\label{36}
\end{equation}
These integrals can be estimated in a long-time regime using the stationary
phase approximation, where the stationary points are approximately determined
by $dE_{k,s}/dk =0$, i.e., $k=0, \pi/2, \pi$. Since our description
involves only $0<k<\pi$, we see that two stationary points lie at the edge of
the interval, and therefore their contribution appears with a factor of $1/2$. On
the other hand, the midpoint $k=\pi/2$ is also the point of maximal approach
between bands, and it only contributes when $\mu\neq 0$. From (\ref{35}) and
(\ref{36}), we see that $x_\zitt$ contains terms with a time dependence of the form
$e^{i \omega t} \sqrt{1/t}$, after applying the stationary phase approximation.
Therefore, the frequencies of oscillation take the values $\omega_1=\pm \mu$
(from $k=\pi/2$) and $\omega_2=\pm \sqrt{4\Delta^2 + \mu^2}$ (from $k=0,\pi$), while the
effect vanishes with an envelope curve $\sqrt{1/t}$. We have an expression of the form
\bea
x_\zitt \approx \sqrt{\frac{1}{t}} \left[ A(\mu,\Delta) e^{-2i \omega_1 t } + B(\mu,\Delta) e^{-2i \omega_2 t } + \mbox{c.c.}\right],\nonumber \\
\label{36.1}
\eea
where $A$ and $B$ are coefficients related to second derivatives of the phase in (\ref{35}) and (\ref{36}). In Fig. \ref{zittercompleta}, we describe the oscillations of $x_\zitt$ in log scale, showing clearly an envelope $\sqrt{1/t}$ for long times.

%{\bf A.R. Fig 10 zittercompleta.eps}
%{\cp I believe that we can give the explicit form of $x_\zitt$, together with 
%the conditions for its validity.}

 % }}}

\begin{figure*} % {{{Evolucion
\includegraphics[width=17cm]{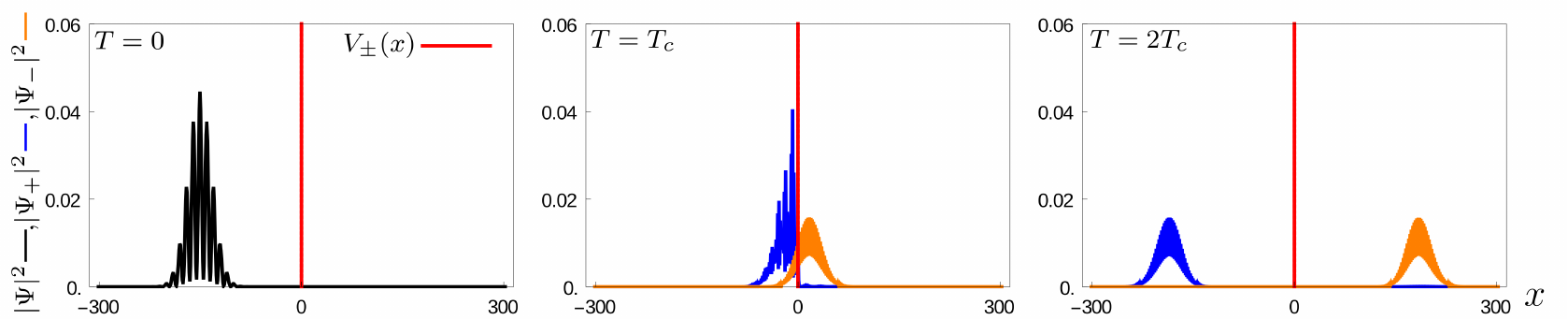} 
\caption{\label{Evolucion} (Color online) Evolution of a wave packet going through the lattice
polarizer. The first picture shows the initial condition of the complete wave packet, whereas
the second and third pictures portray the dynamics of the upper and lower band components
of the wave packet. The collision time with the polarizer is
$T_{c}=\frac{\hbar N}{2 \Delta \kappa}$, where $N$ is the number of sites on
the lattice. }
\end{figure*} % }}}

% }}}
\subsection{The potential as a beam splitter  \label{sec:3B}} % {{{

\begin{figure} % polarizercapacities {{{
\includegraphics[width=\columnwidth]{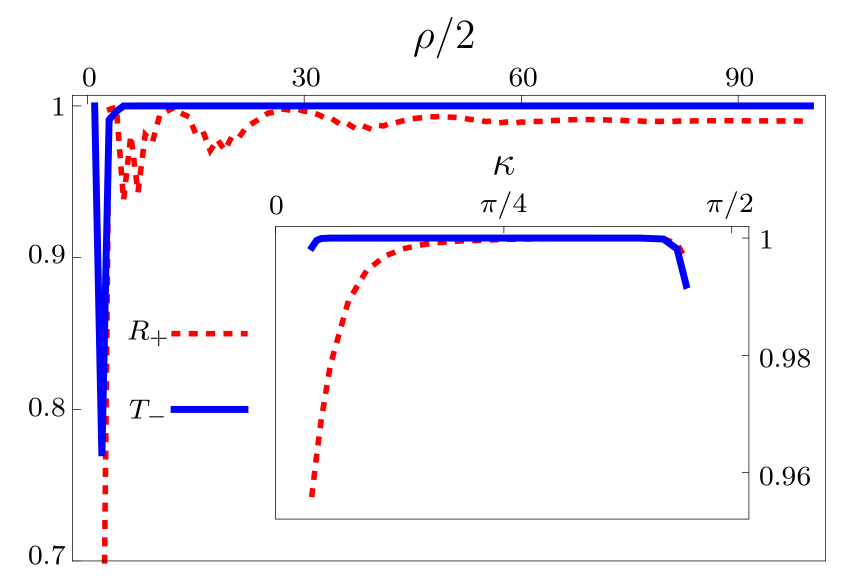} 
\caption{\label{polarizercapacities}(Color online) The dotted line represents
the reflection coefficient for the upper band component of the wave packet as a
function of the variables $\kappa$ and the polarizer size $\rho$. The
continuous line represents the transmission coefficient for the lower band
component of the wave packet as a function of the same variables. }
\end{figure} % }}}

We prepare wave packets with an adjustable width and a proper "thrust" or "kick"
by means of an additional plane-wave factor, imprinting an average drift. Our
choice corresponds to motion from left to right. Eventually, our packets reach
the gate keeper centered at the origin described in Fig.
\ref{polarizermatrix}, but before they do so, \textit{Zitterbewegung} is significantly
observed. After the packets collide with the potential, positive-energy
components are reflected and negative-energy components are transmitted.  This
type of behavior has been verified numerically with specific wave packets, as
we discuss now. In Fig. \ref{Evolucion}, we plot the full probability density
in black, upper spin component in blue, and lower spin in orange. The dynamics
is described in three steps: the first column corresponds to times before the
collision with the polarizer, the second column shows the interference produced
by the collision, and the third column finally demonstrates how the components of
the wave packet are separated after the collision. Upper spin is reflected and
lower spin is transmitted.  To make a quantitative analysis in terms of
probabilities, first we define the initial wave packet as
\begin{equation}
\psi_n(0) =
  \alpha \ncal_- P_{-}e^{-an^2/4 \lambda^2} e^{i \kappa n}+\beta \ncal_+ P_{+}e^{-an^2/4 \lambda^2} e^{-i \kappa n},
\label{37}
\end{equation}
where $|\alpha|^2+|\beta|^2=1$, $\lambda$ is the width of the discrete probability density, and $\kappa$
is the average momentum of the packet. $P_{\pm}$ are the projectors onto each energy band given by
\begin{equation}
P_{s}=\int_0^\pi dk|k,s\rangle\langle k,s|=U_{FW}\left(\frac{\v 1 + s\sigma_{3}}{2}\right)U_{FW}^\dagger.
% P_{\pm}=\sum_{i}|E_{\pm}^{(i)}\rangle\langle E_{\pm}^{(i)}|=U_{FW}\left(\frac{\v 1 \pm\sigma_{3}}{2}\right)U_{FW}^\dagger.
\label{38}
\end{equation}

The matrix elements of these projectors are used after the scattering event takes place in the simulation, in order to test
the sign of the spin. The results in Fig. \ref{polarizercapacities} show that after our Stern-Gerlach
apparatus has done its job, only $1.2\%$ of the upper spin component 
and $100\%$ of the lower spin component have been transmitted. The wave 
packet moving to the right still exhibits a slight hint of \textit{Zitterbewegung} as it is a mixture of components, while
the wave packet moving to the left propagates without \textit{Zitterbewegung}, 
as it is only comprised by the remainder $98.8\%$ of the upper spin component. This quantitative analysis requires
the reflection capacity of the upper spin component, denoted by $R_+$,
and transmission capacity of the lower spin component, $T_-$, of the polarizer
for different values of the thrust $\kappa$ and the range of the polarizer $\rho$. These quantities are given by
% \begin{equation}
%  R_{\pm}=\frac{|P_{\pm}|\psi(t)\rangle|_{l}^2}{|\alpha|^2},T_{\pm}=\frac{|P_{\pm}|\psi(t)\rangle|_{r}^2}{|\alpha|^2},
%  \end{equation}
\begin{multline}
 R_{+}=\frac{|P_{+}|\psi(t)\rangle|_{l}^2}{|\alpha|^2},\quad T_{+}=\frac{|P_{+}|\psi(t)\rangle|_{r}^2}{|\alpha|^2},\\
 R_{-}=\frac{|P_{-}|\psi(t)\rangle|_{l}^2}{|\beta|^2},\quad T_{-}=\frac{|P_{-}|\psi(t)\rangle|_{r}^2}{|\beta|^2},
 \end{multline}
where subscripts $l,r$ stand for sums over sites to the left and right of the
polarizer location, respectively. Due to complementarity, $R_{+}+T_{+}=1$ and
$R_{-}+T_{-}=1$, so $T_{+}$ and $R_-$ are redundant. The results for $R_+, T_-$
are shown in Fig. \ref{polarizercapacities}. When $\rho$ is varied, both
capacities retain near optimal values and fall to zero only for small polarizer
sizes, as expected. Since the "kick" is a property of the wave packet---i.e.,
external to the structure of the polarizer---the capacities are expected to
remain invariant for different values of $\kappa$. This is confirmed in our
simulations, except for values near $\kappa=0, \pi/2$ which correspond to
purely diffusive propagation. 

The results are quite satisfactory, but we should mention that the type of
polarizer $\v (1+\sigma_3) \otimes V$ could be modified with more refined
constructions, even with transparent potentials previously designed using
supersymmetric methods \cite{sadurni2014}.

We would like to point out that the inset in Fig.~\ref{polarizercapacities}
shows the reflection $R_+$ rising up very close to 1 for values of $\kappa >
\pi/4$ (but far from $\pi/2$). This corresponds to fast wave packets. Since our
simulations consist of time-dependent scattering, we need fast and broad
distributions that overcome the spreading of components before scattering; we
are, however, limited to a finite size of the grid. In addition, our model also
allows one to increase the intensity $V$, which blocks incident beams with increasing
efficiency as long as $\kappa$ does not correspond to a Ramsauer resonance.

%{\bf A.R. Discuss robustness of separation of components for
%variations of $\rho$ and $\kappa$ i.e. Fig 9 RTvsrhokappatimes.eps }
% }}}
\subsection{A purely geometric beam splitter  \label{sec:3C}} % {{{

Engineering beam splitters by means of nonlocal potentials include the
possibility of removing all diagonal contributions in $V$, in favor of the
off-diagonal elements representing interactions to a certain range, as shown in Fig. \ref{polarizermatrix} (our
approximations may include nearest neighbors, next-to-nearest neighbors, and so
on). In the experimental setup to be described in later sections, the
couplings can be determined by proximity between sites. With this technique, we
can control the interaction range, as well as the zone where it operates, only
using lattice deformations. Wave-packet evolution is studied numerically in this
extreme situation and our results show a surprisingly efficient separation of
components. In particular, for a $\rho=10$ polarizer with couplings to second-order neighbors, we see a reflection of $67.9\%$ of 
the upper spin component and a transmission of $92.6\%$ of the lower spin component.

% }}}
% }}}
\section{Feasibility  \label{sec:4}} % {{{

\begin{figure} % {{{
\includegraphics[width=\columnwidth]{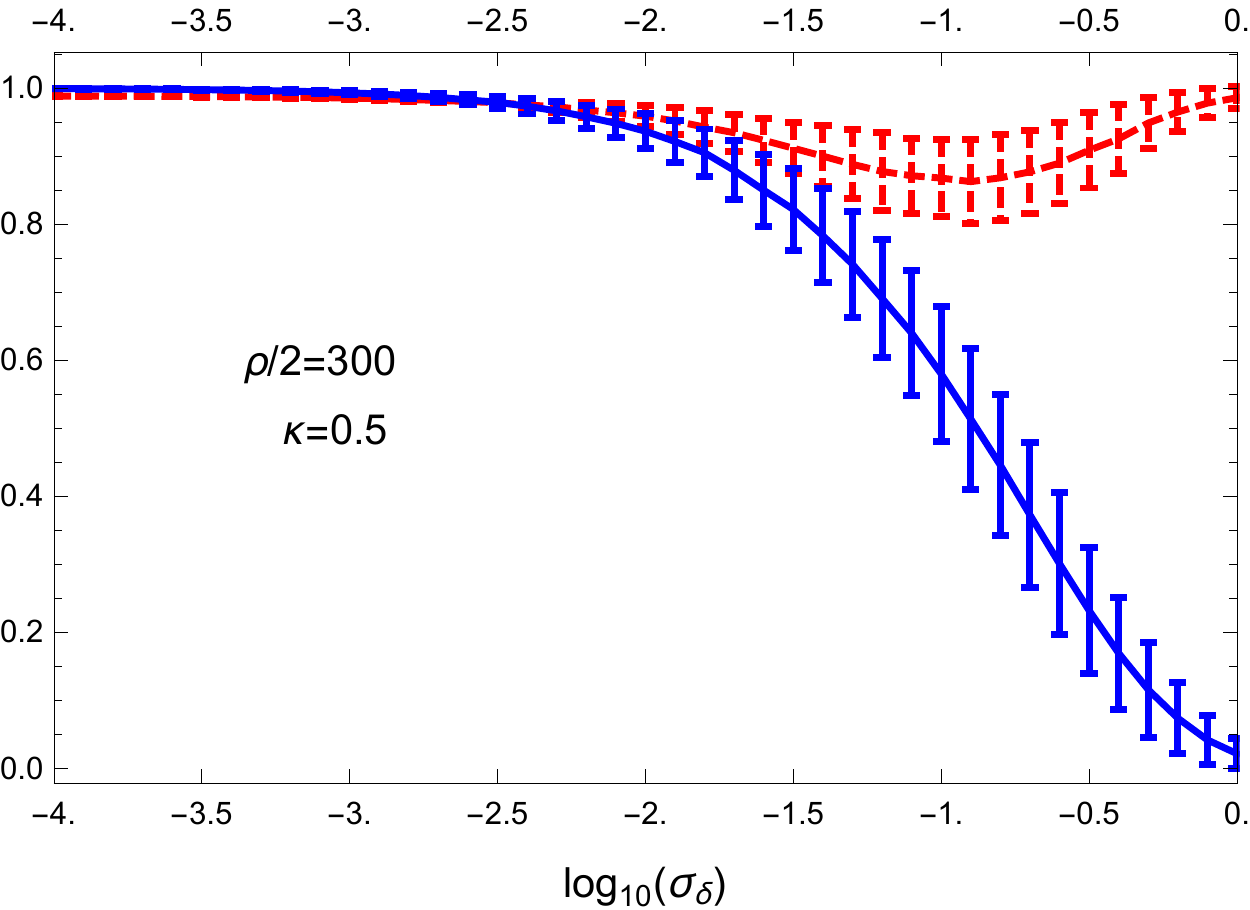}
\caption{\label{fig:feasibility} 
(Color online) Mean reflection (red) and transmission (blue) coefficients as functions of the standard deviation of the coupling $\sigma_\delta$ (see text), for a $\rho=600$ splitter. The error bars represent the fluctuations obtained from multiple realizations. The value $\kappa=0.5$ has been chosen for optimality.}
\end{figure} % }}}

In this section, we test the robustness of the splitter with respect to the known experimental limitations. In the splitter, three parameters must be controlled: the overall absorption, the on-site energy, and the coupling terms.
This analysis will not include the overall absorption because it mainly affects the width and height of the resonances without significantly disturbing the spectral positions; therefore, it is expected that the transmission and reflection coefficients decrease by a factor related to the strength of the absorption.\\
Experiments show \cite{bellec2013} that the on-site energy can be controlled better than the coupling. This is the case for microwave experiments since the variation of couplings is at least two orders of magnitude greater than the variation of the on-site energy. Thus, to estimate the robustness of the splitter, we will consider a Gaussian disorder introduced randomly on the couplings. We modify $\Delta\rightarrow(1-\delta)\Delta$, where $\delta$ is a random variable with a standard deviation $\sigma_\delta$.
Figure \ref{fig:feasibility} shows that the expected coefficients and deviations are satisfactory, even for a poor coupling control ($\sigma_\delta\sim0.1$). As expected, an extremely poor coupling control ($\sigma_\delta\gg0.1$) destroys the efficiency of the splitter, with the latter becoming a regular wall unable to separate the upper and lower band components. Thus, we have shown robustness and feasibility in laboratory implementations.

% }}}
\section{Experimental proposals  \label{sec:5}} % {{{

In this section, we describe a realization of the splitter through
a microwave cavity containing a set of cylindrical resonators between parallel plates, establishing a tight-binding configuration. 
This type of experimental implementation has been very useful for the emulation 
of Dirac equations~\cite{franco2013}, graphenelike structures~\cite{bittner2010,kuhl2010,barkhofen2013,bellec2013}, chiral states~\cite{dembowski2003}, and anomalous Anderson localization~\cite{fernandez2014}, among others. It is important to mention that the following experimental proposal is not unique since the splitter can also be achieved by plasmonic circuits~\cite{2014NatSR...4E7314M}, optical waveguides~\cite{Longhi2009}, or acoustic waves~\cite{Maynard2001}.
The reader can notice that these implementations rely on classical aspects of
the systems mentioned. However, the equations of motion are equivalent to, say,
the Schrödinger or Dirac's equation, depending on the regime studied. In this sense, 
we are {\it emulating} Dirac's equation.

%In this section we show experimentally that the building blocks of our arrays
%are realistic by producing them in microwave cavities with ceramic resonators,
%playing the role of atomic sites (individual resonances are the analogues of
%atomic orbitals). 

We show in further detail how to produce complex coupling
constants with the aim of fabricating purely geometric beam splitters. The
effect, important in its own right, rests on the possibility of breaking the
chiral symmetry of polygonal geometries using dimers as individual sites. This
opens the possibility of producing directed couplings, emerging from dimeric
states. 

\subsection{Experimental specifications  \label{sec:5A}} % {{{

A set of cylindrical dielectric disks can act as the sites of the chain, 
for example, Temex-Ceramics disks, E2000 series, with high dielectric permittivity ($\epsilon=37$)
and low loss (quality factor $Q=7000$). Each disk has an isolated resonance defined by the 
dimensions of the cylinder, e.g., for a height of $5$mm and a radius of $4$mm, a resonance close 
to 6.64 GHz appears corresponding to the lowest transverse electric mode (TE$_1$).
This resonant frequency is equivalent to the on-site energy. 
For purely geometric splitters, we have seen that on-site energies are the same throughout the array; 
therefore, identical dielectric disks must be used. On the other hand, a general type of splitter would require 
disks of different dimensions and/or dielectric constants.

Between two parallel metallic plates, each isolated resonance behaves like a $J_0$-Bessel function 
inside of a cylinder, and as a $K_0$-Bessel function outside of it. The function $K_0$ 
can be represented fairly well by an exponential tail as a function of the distance with respect to the center. Therefore, any set of disks interacts by proximity through the overlap of 
their individual functions $K_0$, in such a way that the response of the whole set is well 
described by a tight-binding model.
The intensity of the interaction and the main contribution of first and second neighbors 
can be further manipulated by changing the distance between the plates~\cite{sadurni2013}.

It is possible to study the wave dynamics of the splitter by introducing two antennas into the microwave cavity connected to different ports of a vector network analyzer (VNA). It is possible
to measure both the spectrum and the intensity of the wave functions by using only one probing antenna. However, for the reconstruction of wave packet dynamics, it
is necessary not only to measure the intensity but also the phase. Hence a second antenna probing the transmission 
of the system is mandatory.

We fix one of the antennas near to a disk whereby the electromagnetic waves are injected, while the position of the other antenna is varied throughout 
the structure, allowing one to measure the transmission spectrum on each disk.

The evolution of the wave packet at each point of the structure is reconstructed through a Fourier transform 
of the measured spectrum at that point~\cite{bohm2015}. This is allowed because we have access to the full spectrum of the complex transmission.

% }}}
\subsection{Negative couplings and level inversion  \label{sec:5B}} % {{{

\begin{figure} % {{{
\includegraphics{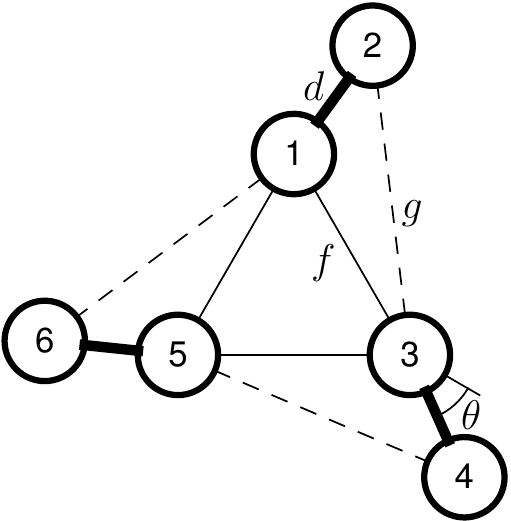}
\caption{\label{fig:examples:cthree:symmetry}
Configuration of disks (indicated as circles, with a number) giving rise to the 
coupling structure specified by Eq.~(\ref{eq:interaction:dimers}). The six disks
are organized in pairs $(1,2)$, $(3,4)$, and $(5,6)$, each of which interacts strongly 
via the coupling constant $d$. The inner disks interact via the coupling constant
$f$. The outer disks interact with just one of the other four disks (for example, 2 with 
3), as the others remain screened geometrically. The $C_{3v}$ symmetry is broken 
by tilting the outer disks with an angle $\theta$. }
\end{figure} % }}}
\begin{figure} % {{{
\includegraphics{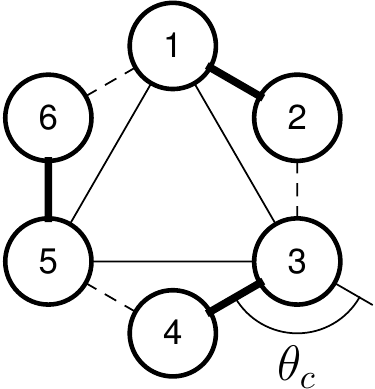} \hfill \includegraphics{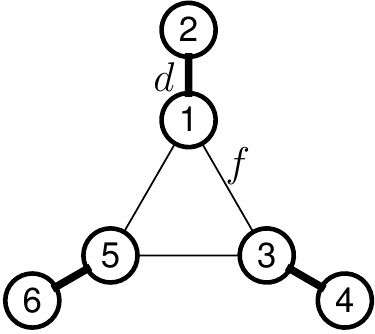}
\caption{\label{fig:examples:cthree:particulars} Two particular realizations
that keep the full $C_{3v}$ symmetry are illustrated, one in which $\theta = \theta_c$, 
and the other for $\theta=0$ in which screening sets $g=0$. 
%{\cp Ojo, $\theta_c$ no 
%siempre es $2\pi/3$. Creo que $\cos \theta = D/2R$ con $D$ la distancia del centro del 
%triangulo al centro de 1 y $R$ la distancia del centro de 1 al centro de 2.}
}
\end{figure} % % }}}

\begin{figure} % {{{
\includegraphics[width=\columnwidth]{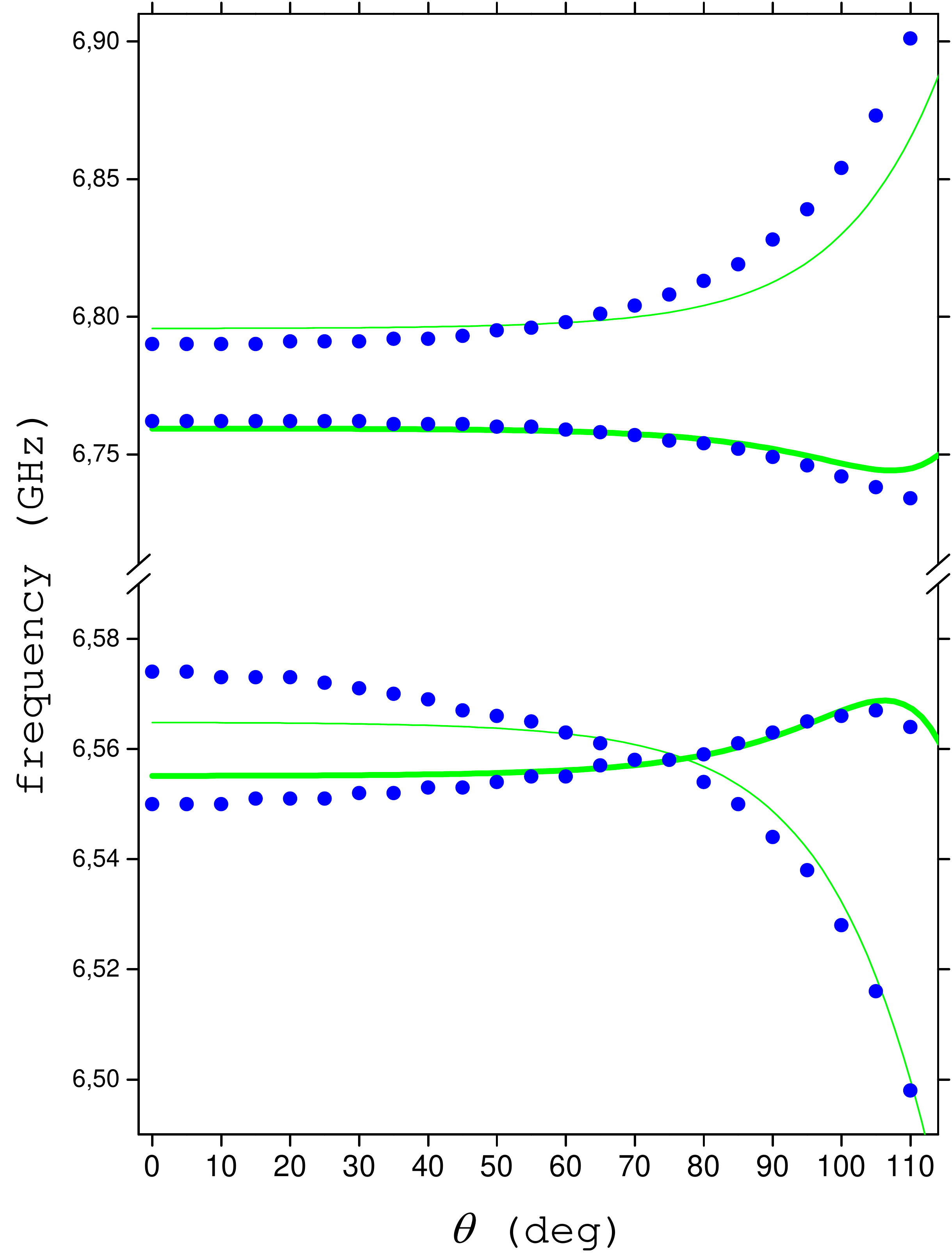}
\caption{\label{fig:1.2} Spectrum of the configuration shown in Fig.~\ref{fig:examples:cthree:symmetry}; dots correspond to a full 3D simulation of a microwave cavity using COMSOL 5.2 and continuous lines correspond to tight-binding calculations. The lower band shows the desired inversion level due to effective negative coupling.}
\end{figure} % }}}

\begin{figure} % {{{
\includegraphics[width=\columnwidth]{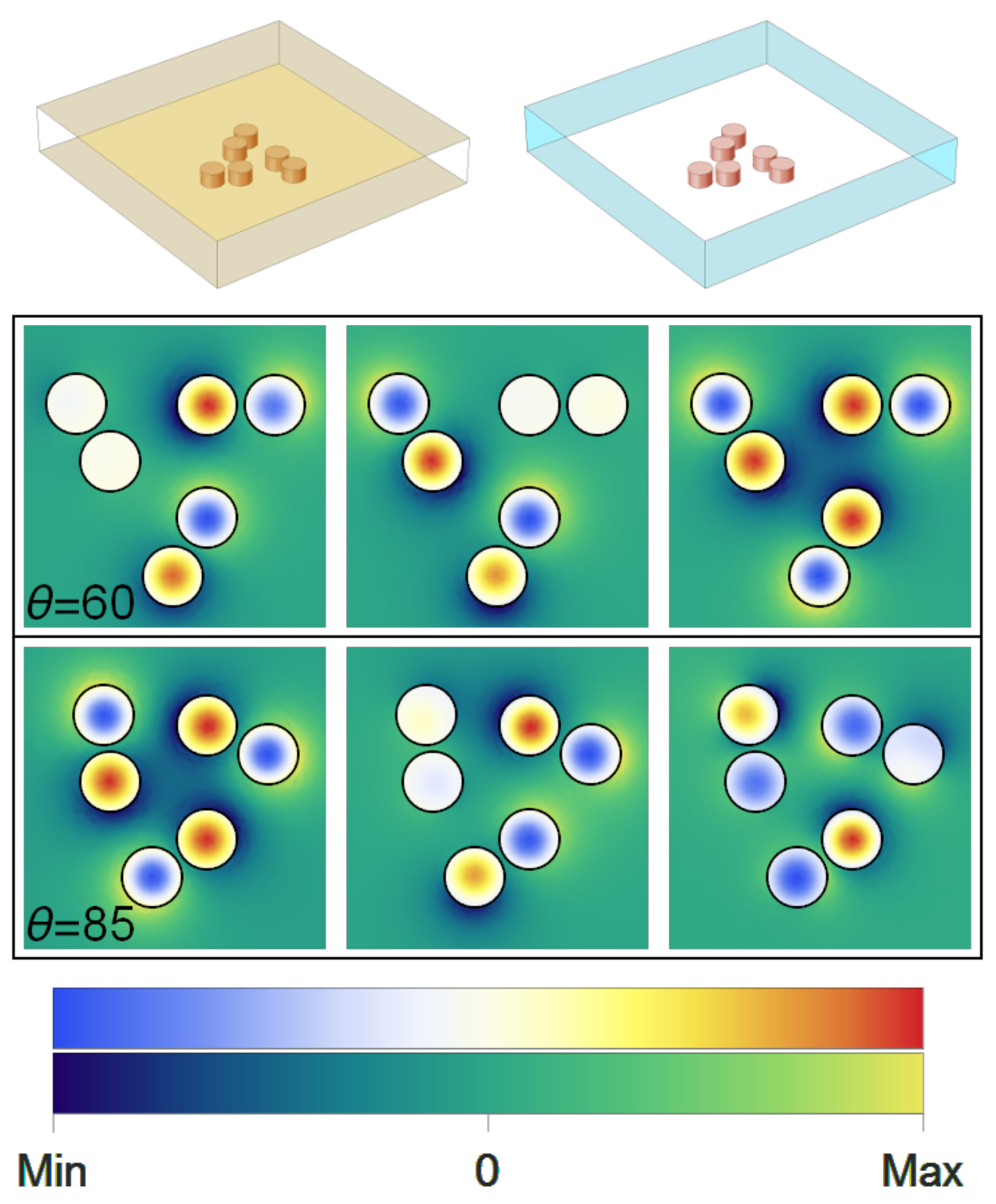}
\caption{\label{fig:modes} Upper row: simulated 3D system; dielectric disks depicted in light brown, nonreflective walls in cyan, and perfect conductors in yellow. Lower row: lowest modes for $\theta=60$ (before level inversion) and $\theta=85$ (after level inversion). Two different scales have been used inside and outside the cylinders for better visibility. The wave functions outside the cylinders exhibit the nature of couplings.}
\end{figure} % }}}

In our purely geometric splitter, we find matrix elements that are real but not
positive; see, e.g., Fig. \ref{polarizermatrix}. 
% According to our previous discussion,
% generating positive couplings entails the variation of distances between
% resonant sites.
% {\cp decir aca que en principio no se puede corregir eso, si es que lo podemos
% comprobar, o si no lo podemos comprobar, que preferimos desarrollar una forma
% mas flexible en el experimento}
Negative couplings
% on the other hand,
require the control of an extra degree of freedom in the
form of a phase factor.
We show that indeed such phases can be produced by
adding more structure in our arrays. It is worth mentioning that nonremovable
phase factors in hopping amplitudes are the equivalent of magnetic fields
applied to charged particles \cite{hofstadter1976}, but our goal is to emulate these effects for a scalar wave. 

First we note that any Hermitian matrix $H$ can be rewritten as a matrix with
semipositive secondary diagonals by means of a unitary transformation. 
% The
% matrix representing our purely geometric polarizer in fig.~\ref{} falls into
% this class {\cp pues es hermitica, entonces si\ldots}, and 
We proceed to turn $H$ into a purely positive nearest-neighbor
array.  Consider
\begin{equation}
U_{\mbox{\scriptsize sign}} = \mbox{diag} \left\{ e^{-i \Delta_n} \right\},
% U_{\mbox{\scriptsize sign}} = \mbox{diag} \left\{ e^{i \delta_n} \right\}.
\label{n1}
\end{equation}
where $\Delta_n = \sum_{m<n} \arg H_{m+1,m}$ is the accumulated phase of the
elements in the first diagonal.
This trivial "gauge" transformation moves all possible phases to third
diagonals or next-to-nearest neighbors; we must now analyze the influence of
sign flips in the hopping amplitudes. The zigzag arrays shown in Fig. \ref{fig:-0.1}
are made of alternating triangular blocks; therefore, every negative sign
occurring in our polarizer corresponds to those bonds lying on the outer part
of the array (see Fig. \ref{fig:-0.1}). For this reason, we focus on a single
triangular block. 

The effect of a negative matrix element in this case produces
level inversion, as shown by the Hamiltonians
\begin{equation}
H^{\pm}_{\mbox{\scriptsize Block}} = \left( \begin{array}{ccc}  E_0 & \Delta & \pm \Delta  \\  \Delta  & E_0 & \Delta \\  \pm \Delta & \Delta & E_0 \end{array} \right)
\label{n2}
\end{equation}
which are related by the unitary transformation $U_{\mbox{\scriptsize Block}} =
\mbox{diag} \left\{ -1,1,-1 \right\}$ in the form
\begin{equation}
U_{\mbox{\scriptsize Block}} H^{-}_{\mbox{\scriptsize Block}}  U_{\mbox{\scriptsize Block}}^{\dagger} = 2E_0 - H^{+}_{\mbox{\scriptsize Block}}. 
\label{n3}
\end{equation}
This compels one to consider each triangular block on the polarizer
as a level-inverting interaction. The simplest way to produce a level-inverted
band is by the introduction of dimers instead of single-resonance sites; see
Fig.~\ref{fig:examples:cthree:symmetry}. 

In the ideal situation where only a change of sign is intended, the dimers are placed such that the $C_3$ symmetry of the array is not destroyed.
To this end, the orientation of the
dimers must be constrained, as shown in Fig. \ref{fig:examples:cthree:particulars}. Note, however, that the full
symmetry of an equilateral triangle $C_{3v}$ is now, in general, broken.
The resulting
shapes are hexagonal variants described by the following tight-binding matrix:
\begin{equation}
\left(\begin{array}{cccccc}0& d& f& 0& f& g\\ d& 0& g& 0& 0& 0\\ f& g& 0& d& f& 0 \\ 0& 0& d&
   0& g& 0 \\ f& 0& f& g& 0& d \\ g& 0& 0& 0& d& 0 \end{array} \right).
\label{eq:interaction:dimers}
\end{equation}
The spectrum contains two degenerate doublets and two singlets. Moreover, their
eigenfrequencies are symmetrically disposed around $E_0$. In essence, we
have produced an additional inverted copy of the spectrum due to
a splitting caused by strong intradimer coupling. For dielectric disks, a numerical simulation of Maxwell equations with space-dependent dielectric functions has been run. The results in Fig. \ref{fig:1.2} show that
the inverted copy corresponds to eigenfrequencies sitting to the left of the
original isolated resonance at $E_0$. Moreover, this occurs only for $\theta > \theta_c \sim 78$ deg, which establishes the existence of a diabolic (crossing) point in the spectrum \cite{berry1984}. Transverse modes are shown in Fig. \ref{fig:modes}, where the panels exhibit a change in the sign of the wave function inside at least one dimer, due to the transition at $\theta_c$.

Finally, our results show that the assembled structure of
alternating triangles must produce two bands opening around each level of a
single dimer: we may choose to work in one or the other. 
A similar spectral structure has been achieved in other contexts: nuclear resonances~\cite{brentano1999}, flat microwave cavities~\cite{dembowski2001,bittner2014}, and electronic circuits~\cite{stehmann2004}.

% }}}

% }}}
\section{Conclusion and outlook \label{sec:6}} % {{{

In this paper, we have studied a tight-binding model that is described by a
Dirac equation. We have focused on the time-dependent dynamics in the positive-
and negative-energy bands. In the language of the Dirac equation, this corresponds
to particles and antiparticles. We have developed the theory that allows one to
split these components by means of a localized potential; this in turn could be
a first step towards the actual measurement of quasispin using wave packets.
We have further shown that even though the interactions are long ranged, taking
as few as next-to-nearest-neighbor interactions, in a very localized region in
space, yields reasonable results.  
    In connection with the possibility of generating pure spin waves with our
    splitter, we would like to add that waves with vanishing average momentum
    have been achieved and that quasispin can be indeed spatially transported.
    However, the mechanism relies on deformations rather than the application
    of external magnetic fields as in the usual case of spin.
The local nature of the interaction is
highly desirable if an experimental emulation is pursued. We have indeed
explored such scenario in the context of a bidimensional array of dielectrics
in a microwave cavity. In such an array, it has been necessary to consider level
inversion, which we have demonstrated using a simple geometric array. The next
obvious step would be to carry out the experiment. 

%We have shown that current techniques would allow this experiment to be carried
%out in the energy domain.  An experiment in time domain, where a direct
%demonstration would be realized, require a similar approach. 

% }}}
\begin{acknowledgments} % {{{
Financial support from CONACyT under Projects CB No. 2012-180585 and No. 153190 and 
UNAM-PAPIIT IN111015 is acknowledged. We are grateful to LNS-BUAP for allowing extensive use of their supercomputing facility.
\end{acknowledgments} % }}}
\appendix*
\section{An alternative splitter \label{sec:appendix}} % {{{

A simple alternative splitter can be designed if we replace Eq.~(\ref{17.1}) by 
% The simplest way to separate both types of waves is by introducing a potential of the type
\begin{equation}
V_{\pm}(x) = \begin{cases}
V(x) & \text{for particles} \\
-V(x) & \text{for holes}
\end{cases}.
\end{equation}
Then, Hamiltonian (\ref{18}) would be replaced by 
\begin{equation}
\tilde H_{\mbox{\scriptsize FW}} = H_{\mbox{\scriptsize FW}} + \sigma_3 \otimes  V_{\mbox{\scriptsize FW}}.
\end{equation}
In formula (\ref{v11}), one would need an extra term, 
\begin{multline}
V_{11} = e^{-i\phi/2}\cos\left(\frac{\theta}{2} \right) 
              V_{\mbox{\scriptsize FW}} \cos\left(\frac{\theta}{2} \right) e^{i\phi/2} \nonumber \\
+ e^{-i\phi/2}\sin\left(\frac{\theta}{2} \right) 
              V_{\mbox{\scriptsize FW}} \sin\left(\frac{\theta}{2} \right) e^{i\phi/2},
% \label{21}
\end{multline}
which leads to the following changes in the matrix elements:
\begin{multline}
\<n|V_{11}|n'\> = 
  \frac{1}{8\pi^2} \sum_{m=-\infty}^{\infty} V_{\mbox{\scriptsize FW}}(m) \\ 
  \times 
  \left[ I_{n'-m}^{++} \left( I_{n-m}^{++}\right)^* 
              + I_{n'-m}^{--} \left( I_{n-m}^{--}\right)^*  
  \right],
% \label{23}
\end{multline}
for even $n$ and $n'$, 
\begin{multline}
\<n|V_{21}|n'\> = 
  \frac{1}{8\pi^2} \sum_{m=-\infty}^{\infty} V_{\mbox{\scriptsize FW}}(m) \\ 
  \times 
  \left[ I_{n'-m}^{--} \left( I_{n-m}^{+-}\right)^* 
      - I_{n'-m}^{++} \left( I_{n-m}^{-+}\right)^*  \right], 
% \label{24}
\end{multline}
for even $n$ and odd $n'$, and, finally,
\begin{multline}
\<n|V_{22}|n'\> = 
  \frac{1}{8\pi^2} \sum_{m=-\infty}^{\infty} V_{\mbox{\scriptsize FW}}(m) \\ 
  \times 
  \left[ I_{n'-m}^{-+} \left( I_{n-m}^{-+}\right)^* 
  + I_{n'-m}^{+-} \left( I_{n-m}^{+-}\right)^*  \right], 
% \label{eq:blocks:odd:odd}
\end{multline}
for odd $n$ and $n'$.

% }}}
% Bibliografia {{{
%\nocite{*}

\bibliography{bibliosplitter}

% }}}
\end{document}
% ****** End of file  ******